\documentclass[bibnotes,twocolumn,aps,prb,a4paper]{revtex4}

\usepackage{color,epsfig,rotating,graphicx,subfigure}
\usepackage{amsmath,amssymb,amscd}

\usepackage[latin1]{inputenc}
\usepackage[english]{babel}

\usepackage{dsfont}
\usepackage{textcomp}

\renewcommand{\Re}{{\rm Re}}
\renewcommand{\Im}{{\rm Im}}

\newcommand{\ri}{{\rm i}}
\newcommand{\re}{{\rm e}}
\newcommand{\rd}{{\rm d}}
\newcommand{\rr}{{\rm r}}
\newcommand{\rt}{{\rm t}}
\newcommand{\rL}{{\rm L}}
\newcommand{\rl}{{\rm l}}

\newcommand{\rs}{{\rm s}}
\newcommand{\rp}{{\rm p}}  

\newcommand{\kb}{k_{\rm B}}

\newcommand{\Tr}{{\rm Tr}}

\newcommand{\rth}{{\rm th}}

\newcommand{\rE}{{\rm E}}

\newcommand{\rB}{{\rm B}}
\newcommand{\rP}{{\rm P}}

\begin{document}

%
%
\title{Near-field heat transfer between a nanoparticle and a rough surface}

\author{S.-A. Biehs and J.-J. Greffet}

\affiliation{Laboratoire Charles Fabry, Institut d'Optique, CNRS, Universit\'{e} Paris-Sud, Campus
Polytechnique, RD128, 91127 Palaiseau cedex, France}

\date{19.03.2009}
\begin{abstract}
In this work we focus on the surface roughness correction to the near-field radiative heat transfer
between a nanoparticle and a material with a rough surface utilizing a direct perturbation theory
up to second order in the surface profile. We discuss the different distance regimes for the 
local density of states above the rough material and the heat flux analytically and numerically. 
We show that the 
heat transfer rate is larger than that corresponding to a flat surface at short distances. At larger distances
it can become smaller due to surface polariton scattering
by the rough surface. For distances much smaller than the correlation length of the surface profile,
we show that the results converge to a proximity approximation, whereas in the opposite limit the rough surface
can be replaced by an equivalent surface layer.
\end{abstract}

\maketitle

\newpage

%
%

\section{Introduction}

Recently, the near-field radiative heat transfer has attracted a lot of theoretical 
and experimental attention~\cite{JoulainEtAl2005,VolokitinPersson2007,VinogradovDorofeyev2009}. 
It was predicted theoretically~\cite{PvH1971} and shown experimentally~\cite{HuEtAl2008,NarayaEtAl2008,RousseauEtAl2009,ShenEtAl2008} that 
the heat flux for distances much smaller than the thermal wavelength $\lambda_\rth = \hbar c / (\kb T )$ can be much greater than that  
predicted by Planck's law, where $\hbar$ is Planck's constant, $\kb$ is Boltzmann's constant, $c$ is the velocity of light in vacuum and $T$ the
temperature. This unusual property might for example be exploited  
for thermo-photovoltaics~\cite{MatteoEtAl2001,NarayanaswamyChen2003,LarocheEtAl2006,FrancoeurEtAl2008} 
and near-field scanning thermal microscopy~\cite{WischnathEtAl2008,KittelEtAl2008}.  

It is common knowledge that the radiative properties of a material depend not only on the material parameters but also on
the surface roughness~\cite{NietoVesperinas2006}. While the effect on far-field properties has been widely studied~\cite{Maradudin2007},
the impact of surface roughness on near-field heat transfer has not been considered so far.
 From the experimentalist's point of view, at least an estimate of 
the surface roughness correction is desirable, since one is confronted with a certain
degree of surface roughness in all experimental set-ups. 

In this work we will
study the near-field heat transfer between a nanoparticle considered to be an electric dipole and a material with a rough surface.
Within this model the change of the local density of states (LDOS) above the material due to the surface roughness completely causes 
the change in the heat flux. Since, the near-field heat transfer between two semi-infinite bodies is also largely determined by the LDOS,  
we think that our results are not only restricted to the here discussed geometry, but can also be utilized to get a rough estimate 
of the impact of surface roughness for configurations used in recent experimental setups~\cite{HuEtAl2008,NarayaEtAl2008,RousseauEtAl2009,ShenEtAl2008}.
In addition, the LDOS also plays a fundamental role for other physical phenomena as it determines, for instance, the lifetime of atoms 
and molecules near a surface, so that the here given results for the LDOS have a wider range of application.

The paper is organized as follows: In Sec.\ II we give a short description of the dipole model of near-field
heat transfer. In Sec.\ III we introduce the perturbation result for the mean LDOS,
the key quantity for understanding the roughness correction to the mean heat flux which is itself discussed 
in Sec.\ IV. In Sec.\ V we derive approximations for the small and the large distance regime. Finally, in Sec.\ VI we
study the roughness correction to the LDOS numerically.

%
%

\section{Radiative heat transfer}

\begin{figure}[Hhbt]
  \epsfig{file=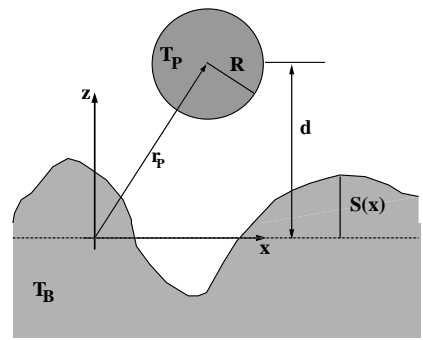, width=0.3\textwidth}
  \caption{\label{Fig:SphereSurface} Sketch of the configuration considered here.}
\end{figure}

We consider the situation depicted in Fig.~\ref{Fig:SphereSurface}. A nanoparticle 
with a polarizability $\alpha$ in local thermal equilibrium at temperature $T_\rP$ 
is placed at $\mathbf{r_\rP}$ near a dielectric half-space with a given permittivity $\epsilon$.
As discussed in Refs.~\cite{MuletEtAl2001,PinceminEtAl1994} the multiple scattering between the nanoparticle
and the surface can be neglected.
We assume that this dielectric is in local thermal equilibrium at a 
temperature $T_\rB \neq T_\rP$. 
The interface separating the dielectric from the vacuum is described
by the surface profile $S(\mathbf{x}) - z = 0$ with $\mathbf{x} = (x,y)$. 

Now, as far as the radius $R$ of the nanoparticle is smaller than the thermal 
wavelength $\lambda_{\rm th}$ and the distance $d$ between the dielectric body and the particle can be assumed to be 
large, i.e., $\lambda_{\rm th} \gg R$ and $d \gg R$, the energy transfer rate 
between the particle and the dielectric body due to radiation can be expressed 
within the dipole model as~(see Refs.~\cite{VolokitinPersson2007,VinogradovDorofeyev2009}
and references therein)
\begin{equation}
 \langle P^{\rP \leftrightarrow \rB} \rangle = \int_{0}^\infty\!\!\!\!\rd \omega\, 
                                               2 \omega \alpha''(\omega)
                                               \bigl[ \Theta(\omega,T_\rP) - \Theta(\omega,T_\rB)]
                                               D^{\rE}(\omega,\mathbf{r}_\rP)
\label{Eq:PheatDipole}
\end{equation}
with
\begin{equation}
  \Theta(\omega,T) = \frac{\hbar \omega}{\re^{\hbar \omega \beta} - 1}
\end{equation}
where $\beta = (\kb T)^{-1}$, $\kb$ is Boltzmann's constant and $(2 \pi \hbar)$ is Planck's constant.
Here, the spectral power absorbed by the nanoparticle~\cite{LandauLifshitz2002,VolokitinPersson2007,VinogradovDorofeyev2009} is given by the term 
$\alpha''(\omega) D^\rE (\omega,\mathbf{r}_\rP) \Theta(\omega,T_\rB)$, i.e., it is proportional
to the imaginary part of the polarizability $\alpha''$ of the particle and proportional to the
electric energy density which is given by the product $D^\rE(\omega,\mathbf{r}_\rP)\Theta(\omega,T_\rB)$ 
of the electric local density of states (LDOS)
above the dielectric medium and the mean energy of a harmonic oscillator $\Theta(\omega, T)$. 
On the other hand the power emitted by the particle
and absorbed within the bulk medium is proportional to $\alpha''(\omega) D^\rE (\omega,\mathbf{r}_\rP) \Theta(\omega,T_\rP)$.
We point out that the above given formula has to be augmented by its magnetic counterpart when considering metallic 
materials as discussed in Ref.~\cite{ChapuisEtAl2008}. 

Within this work we will use the expression of the LDOS 
\begin{equation}
  D^\rE(\omega,\mathbf{r}) = \frac{\omega}{\pi c^2} \Im\, \Tr \,\mathds{G}^\rE(\mathbf{r,r})
\label{Eq:Def_LDOSev}
\end{equation} 
as defined in Ref.~\cite{JoulainEtAl2003}. Using Eq.~(\ref{Eq:Def_LDOSev}) in Eq.~(\ref{Eq:PheatDipole}) corresponds to a situation where the
bulk and its surrounding are at ambient temperature $T_\rB$, whereas the nanoparticle is heated
or cooled with respect to $T_\rB$.

%
%

\section{Local density of states}

\subsection{Stochastic surface roughness}

In this work, we concentrate on the special case of a stochastic surface profile $S$ described 
as a stochastic Gaussian process with mean value and correlation function given by
\begin{align}
  \langle S(\mathbf{x}) \rangle_\rp &= 0, \\
  \langle S(\mathbf{x}) S(\mathbf{x}') \rangle_\rp &= \delta^2 W(|\mathbf{x - x'}|).
\end{align}
The brackets $\langle \circ \rangle_\rp$ stands for the average over an ensemble of realizations of 
the surface profile $S(\mathbf{x})$; $\delta$ is the rms height
of the surface profile. The correlation function $W(|\mathbf{x - x'}|)$ is here assumed to be given
by a Gaussian
\begin{equation}
  W(|\mathbf{x - x'}|) = \re^{-\frac{|\mathbf{x - x'}|^2}{a^2}}
\end{equation}
introducing the transverse correlation length $a$.

For the Fourier component $\tilde{S}(\boldsymbol{\kappa})$ of the surface profile function one obtains 
\begin{align}
  \langle \tilde{S} (\boldsymbol{\kappa}) \rangle_\rp &= 0 \\
  \langle \tilde{S} (\boldsymbol{\kappa}) \tilde{S} (\boldsymbol{\kappa}')\rangle_\rp &= (2 \pi)^2 \delta^2 
                      \delta(\boldsymbol{\kappa} + \boldsymbol{\kappa}') g(\kappa)
\end{align}
with Dirac's delta function $\delta(\boldsymbol{\kappa} + \boldsymbol{\kappa}')$ and the surface roughness power spectrum
\begin{equation}
  g(\kappa) = \int\!\!\rd^2 x\, W(|\mathbf{x}|) \re^{- \ri \boldsymbol{\kappa} \cdot \mathbf{x}} = \pi a^2 \re^{-\frac{\kappa^2 a^2}{4}}.
\end{equation}

\subsection{Perturbation expansion of the LDOS}

In order to determine the perturbation expansion of the LDOS in Eq.~(\ref{Eq:Def_LDOSev}), we expand
 the Green's dyadic $\mathds{G}^{\rE}$ with respect to the surface profile up to second order~\cite{Greffet1988}.
We follow Ref.~\cite{HenkelSandoghdar1998}, where one can find a procedure for determining the first-order Green's dyadic. The explicit second order form is given in appendix A. A detailed discussion of the validity of the perturbation theory is 
given in Ref.~\cite{HenkelSandoghdar1998}.  In summary, the perturbation ansatz is valid as far 
as the surface height $\delta$ is the smallest length-scale of the problem, i.e., $\delta \ll \min\{z,a,\lambda_\rth\}$. 

\begin{figure}[Hhbt]
  \epsfig{file=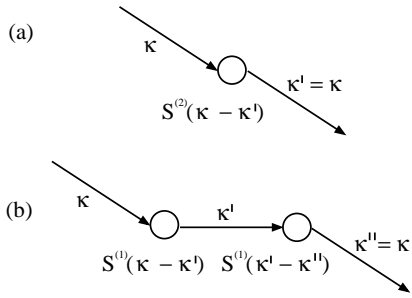, width=0.3\textwidth}
  \caption{\label{Fig:ScatteringProcesses}  Simple sketch of the scattering processes due to $M_{\rs/\rp,0}$ (a) and due to $M_{\rs/\rp,1}$ (b). After averaging  
          the translational symmetry is restored so that $\boldsymbol{\kappa}$ is the same before and after the scattering with the rough surface. Mathematically,
          this property follows directly from the statistical properties of the gaussian surface roughness, i.e.,  
          $\langle \tilde{S}^{(2)} (\boldsymbol{\kappa} - \boldsymbol{\kappa}') \rangle_\rp \propto \delta(\boldsymbol{\kappa} - \boldsymbol{\kappa}')$
          and $\langle \tilde{S}^{(1)} (\boldsymbol{\kappa} - \boldsymbol{\kappa}') \tilde{S}^{(1)} (\boldsymbol{\kappa}'- \boldsymbol{\kappa}'') \rangle_\rp \propto \delta(\boldsymbol{\kappa} - \boldsymbol{\kappa}'')$. 
            }
\end{figure}

Inserting the Green's function up to second order given in Eq.~(\ref{Eq:GreensfunctionAppendix}) into Eq.~(\ref{Eq:Def_LDOSev}),
we find for the LDOS up to second order after ensemble average:
\begin{equation}
\begin{split}
 \langle {D^\rE}^{(0)-(2)}(\omega,\mathbf{r}) \rangle_\rp &= \\
  &\!\!\!\!\!\!\!\!\!\!\!\!\!\!\!\!\!\!\!\!\!\!\!\!\!\!\!\!\!\!\!\!\!\!\!\!\!  \frac{\omega}{\pi^2 c^2} \biggl[
                                           \int_{\kappa \leq k_0}\!\!\rd \kappa\,\frac{\kappa}{4 \gamma_\rr}
                                                \biggl( 1 + h_\rs \Re\bigl(\langle {r_\rs}^{(0)-(2)} \rangle_\rp \re^{2 \ri\gamma_\rr z}\bigr)\biggr)\\
  &\!\!\!\!\!\!\!\!\!\!\!\!\!\!\!\!\!\!\!\!\!\!\!\!\!\!\!\!\!\!\!\!\!\!\!\!+ \int_{\kappa > k_0} \rd \kappa\, \frac{\kappa \re^{-2 \gamma z}}{4 \gamma}
                                         \biggl(h_\rs \Im(\langle {r_\rs}^{(0)-(2)} \rangle_\rp)\biggr) + (\rs \leftrightarrow \rp)\biggr].
\end{split}
\label{Eq:LDOS_sec_ord}
\end{equation}
where $h_\rs = 1$ and $h_\rp = (2 \kappa^2 - k_0^2)/k_0^{2}$.
The first term yields the propagating mode contribution, i.e., for $\kappa \leq k_0$ with 
$\gamma_\rr = \sqrt{k_0^2 - \kappa^2}$ purely real, whereas the second term gives the contribution due to evanescent waves,
i.e., $\kappa > k_0$ with $\gamma_\rr = \ri \gamma$ purely imaginary. The mean reflection coefficient up to second order
$\langle r_{\rs/\rp}^{(0)-(2)} \rangle_\rp$ can be written as a sum of the usual Fresnel coefficient $r_{\rs/\rp}$ and a surface roughness correction
\begin{equation}
   \langle r_{\rs/\rp}^{(0)-(2)} \rangle_\rp = r_{\rs/\rp} - 2 \ri \gamma_\rr (D_{\rs/\rp}^0)^2 M_{\rs/\rp},
\label{Eq:reflspec}
\end{equation}
where
\begin{equation}
  D^0_{\rs} = \frac{\ri}{\gamma_\rr + \gamma_\rt} \qquad \text{and} \qquad
  D^0_{\rp} = \frac{\ri \epsilon}{\gamma_\rr \epsilon + \gamma_\rt}
  \label{Eq:GreenSurfaceWave}
\end{equation}
with $\gamma_\rt = \sqrt{k_0^2 \epsilon - \kappa^2}$. The so-called proper self-energy $M_{\rs/\rp}$ is defined in Appendix~\ref{Append:DefPropSEnergy}.
Furthermore, one can write the expression for the proper self energy $M_{\rs/\rp}$ as a sum of 
two contributions $M_{\rs/\rp,0}$ originating from terms due to a second-order scattering process
and $M_{\rs/\rp,1}$ originating from terms due to two successive first-order scattering processes (see Fig.~\ref{Fig:ScatteringProcesses}).

%
%

\section{Radiative heat transfer between a nanoparticle and a rough surface}

With the above given relations we can study the radiative heat transfer between a spherical nanoparticle and a semi-infinite
dielectric body with a rough surface formally given by Eq.~(\ref{Eq:PheatDipole}) 
setting the nanoparticle at position $\mathbf{r}_\rP = (0,0,d)$. 
For describing the absorptivity of the nanoparticle
with radius $R$ we utilize the polarizability given as
\begin{equation}
  \alpha(\omega) = 4 \pi R^3 \frac{\epsilon(\omega) - 1}{\epsilon(\omega) + 2}. 
\end{equation} 
Here, we employ the material properties for SiC from Ref.~\cite{ShchegrovEtAl2000} for numerical evaluation 
of Eq.~(\ref{Eq:PheatDipole}) using Eq.~(\ref{Eq:LDOS_sec_ord}) and
determine the surface roughness correction defined by
\begin{equation}
  \Delta P = 100\, \frac{\langle P^{(0) - (2)} \rangle_\rp - P^{(0)}}{P^{(0)}}.
\end{equation}
We have checked that $P^{(0)}$ gives the same result as in Ref~\cite{MuletEtAl2001} when choosing the same radius.
(Note that $\Delta P$ itself does not depend on $R$, since $\langle P^{(0) - (2)} \rangle_\rp \propto R^3$ and $P^{(0)} \propto R^3$.)

\begin{figure}[Hhbt]
  \epsfig{file=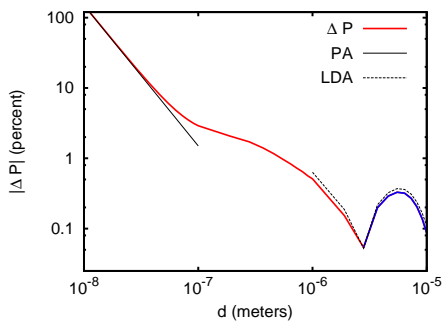, width=0.35\textwidth}
  \caption{\label{Fig:DeltaP} (Color online)
           Plot of the modulus of $\Delta P$ for evanescent modes over the distance for SiC 
           setting $T_\rB = 0\,{\rm K}$ and $T_\rP = 300\,{\rm K}$
           for a rough surface with $\delta = 5\,{\rm nm}$ and $a = 200\,{\rm nm}$. The red part of the curve
           indicates positive and the blue one negative values.
           The PA and the LDA are included for comparison.}
\end{figure}

Let us first turn to the distance dependence of the heat flux. 
A plot of $\Delta P$ over the distance considering only evanescent modes, with $\delta = 5\,{\rm nm}$ and 
$a = 200\,{\rm nm}$ is shown in Fig.~\ref{Fig:DeltaP}.
The temperature of the dielectric is assumed to
be $0\,{\rm K}$, whereas the temperature of the nanoparticle is set to $300\,{\rm K}$. From formula (\ref{Eq:PheatDipole})
it is clear that (apart from a sign for $P$) one gets the same result for $\Delta P$ when interchanging the temperatures.
As will be shown later, the proximity approximation (PA) 
\begin{equation}
  \langle P \rangle_\rP^{\rm PA} = P^{(0)} \biggl(1 + 6\frac{\delta^2}{d^2} \biggr)
\label{Eq:PArht}
\end{equation}
can be derived in the small distance regime with $\delta \ll d \ll a$. Note 
that $\Delta P$ is always positive and independent from $T$ in that limit even if $\epsilon$ 
depends on $T$. In the opposite limit with $d \gg a$, we can also derive a large distance approximation (LDA)
as will be shown later (see Eq.~(\ref{Eq:Mrs1largedist}) and (\ref{Eq:Mrp1largedist})).

\begin{figure}[Hhbt]
  \epsfig{file=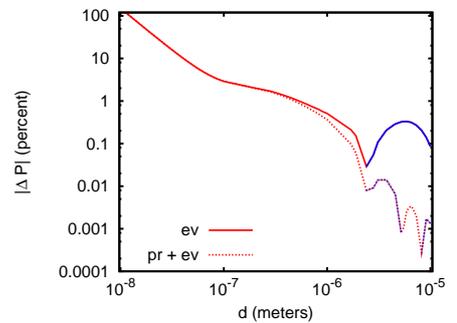, width=0.35\textwidth}
  \caption{\label{Fig:DeltaP_pr_ev} (Color online)
           As Fig.~\ref{Fig:DeltaP} but for evanescent and propagating modes.}
\end{figure}

It can be seen, that $\Delta P$ converges to the approximations for large and small distances. For 
distances slightly greater than the small distance regime well described by the PA result 
the surface roughness correction becomes much greater than predicted by the PA. For distances 
between $1-10\,\mu{\rm m}$ it can be seen that $\Delta P$ becomes negative. Now, in this distance regime already the propagating
modes start to dominate the heat transfer. It turns out that for the propagating modes the surface roughness 
correction is very small compared to that of the evanescent modes. It follows that
$\Delta P$ for evanescent and propagating modes becomes also very small in the distance regime $1-10\,\mu{\rm m}$ as
is illustrated in Fig.~\ref{Fig:DeltaP_pr_ev}. It can also be seen that due to the competition of the roughness correction of the evanescent and propagating
modes, the overall correction to the heat transfer $\Delta P$ becomes more 
oscillatory in that distance regime.
Nevertheless, corrections on the order of ten percent can arise for distances smaller than $100\,{\rm nm}$, but  
one has to keep in mind that the theory used here is only applicable for $d \gg \delta$ and $d \gg R$. Therefore, using $\delta = 5\,{\rm nm}$
and $R = 5\,{\rm nm}$ it can be estimated that the theory is only valid for distances $d$ greater than $\approx 50\,{\rm nm}$.

\begin{figure}[Hhbt]
  \epsfig{file=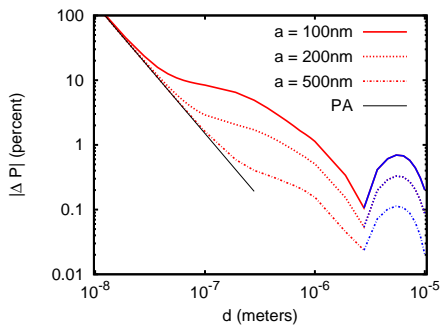, width=0.35\textwidth}
  \caption{\label{Fig:DeltaP_versch_a} (Color online)
           As Fig.~\ref{Fig:DeltaP}
           using $a = 100\,{\rm nm}$, $200\,{\rm nm}$ and $500\,{\rm nm}$.}
\end{figure}

Since we have determined $P$ perturbatively with respect to the surface profile it is clear that to second order $\Delta P$ is proportional
to $\delta^2$. On the other hand, the dependence of $\Delta P$ on the correlation lenght $a$ is not obvious. We find that $\Delta P$ is 
approximately proportional to $a^{-1}$ in the large distance regime (for the here used parameters) what might be seen in Fig.~\ref{Fig:DeltaP_versch_a} considering only evanescent modes
for $a = 100\,{\rm nm}$, $200\,{\rm nm}$ and $500\,{\rm nm}$. 

In order to understand the roughness correction to the heat flux, a deeper understanding of the 
roughness correction to the 
LDOS is required. Therefore,  we will discuss the LDOS in more detail in the following.

%
%

\section{Approximations of the LDOS for small and large distances}

The key quantities for the LDOS are the reflection coefficients in Eq.~(\ref{Eq:reflspec}). Here we will derive
some approximations of the proper self energy $M_{\rs/\rp}$ first. From these
expressions one can get the corresponding approximation for the LDOS
by using Eq.~(\ref{Eq:LDOS_sec_ord}) and Eq.~(\ref{Eq:reflspec}).  

Before we derive the small and large distance approximation for the proper self energy $M_{\rs/\rp}$ we first implement the following
approximation: For frequencies relevant at room temperature, i.e., $\omega \approx 10^{14}\,{\rm s}^{-1}$,
we have $|k_0 \sqrt{\epsilon} a/2| \ll 1$, i.e., $a$ is much smaller than the skindepth $\rd_\rs$, as far as the correlation 
length $a$ is small enough. Considering
SiC, this relation is well fullfilled for most frequencies within the 
range $3.7\cdot10^{13}-2\cdot10^{14}\,{\rm s}^{-1}$ 
for values of $a$ smaller than $500\,{\rm nm}$. Therefore, we concentrate here on the limit $a \ll \rd_\rs$ only (see appendix~\ref{Append:properselfenergyLDA}), while
one can determine the opposite limit $a \gg \rd_\rs$ with a similar procedure. For the latter limit we just state
the results in Appendix~\ref{Append:LargeDistLimit}.

\subsection{small distance regime ($\delta \ll z \ll a$)}

In the limit of small distances $z$ it is seen from Eq.~(\ref{Eq:LDOS_sec_ord}) that the main contribution comes from wavevectors $\kappa \approx 1/z$.
In the regime $a \gg z$ ($\kappa a \gg 1$), the 
terms $M_{\rs/\rp,1}$ become negligible compared to $M_{\rs/\rp,0}$. This is due to the fact that for
$\kappa a \gg 1$ the proper self-energy contributions $|M_{\rs/\rp,1}|$ are proportional to $1/\kappa$ or $\kappa$,
resp., because the Bessel functions are in this limit approximated by $\exp(\xi^2/8)/\xi$. On the other
hand $|M_{\rs/\rp,0}|$ is for large wave vectors proportional to $\kappa$ or $\kappa^3$, so that 
$|M_{\rs/\rp,1}|/|M_{\rs/\rp,0}| \propto 1/\kappa^2$. 
This means, that in the near-field regime the
surface roughness correction solely stems from $M_{\rs/\rp,0}$ in Eq.~(\ref{Eq:Ms0appendix}) and (\ref{Eq:Mp0appendix}), i.e., 
from that term which originates from one second-order scattering process (see also Fig.~\ref{Fig:ScatteringProcesses}). 
Since the term $M_{\rs/\rp,1}$ originating from terms due to two successive first-order scattering processes
including  processes involving the excitation of surface modes  with $(\kappa,\omega)$ followed by scattering into another 
surface mode ($\kappa',\omega$) as an intermediate state which is then scattered back into the initial state 
with ($\kappa,\omega$)
becomes negligible, we can  
formally conclude that these processes are irrelevant for $z \ll a$. This is very intuitive, since the wavelength of the surface modes becomes much smaller than the 
correlation length. Within this limit the excited surface modes can follow the perturbed surface adiabatically 
without being scattered. Hence, for $z \ll a$ it suffices to consider $M_{\rs/\rp,0}$ only.
Inserting $M_{\rs/\rp,0}$ into the mean reflection coefficient in Eq.~(\ref{Eq:reflspec}) gives for the
evanescent near-field regime with $\kappa \gg k_0$
\begin{equation}
  \langle r_{\rs/\rp}^{(0)-(2)}\rangle_\rp \approx \tilde{r}_{\rs/\rp}\bigl(1 + 2 (\kappa\delta)^2 \bigr),
\label{Eq:MeanReflSmallDist}
\end{equation}
for $\kappa a \gg 1$ or $z \ll a$, utilizing the quasistatic approximations for the reflection coefficients
\begin{equation}
  \tilde{r}_\rs = \frac{k_0^2}{4 \kappa^2} (\epsilon - 1) \qquad\text{and} \qquad 
  \tilde{r}_\rp = \frac{\epsilon - 1}{\epsilon + 1}.
\label{Eq:FresnelQuasistatic}
\end{equation}
From the above relations we can infer that the surface roughness correction for very small distances does not depend on $a$.
The expressions in Eq.~(\ref{Eq:MeanReflSmallDist}) can also be derived for $a \gg \rd_\rs$.

By inserting Eqs.~(\ref{Eq:MeanReflSmallDist}) and (\ref{Eq:FresnelQuasistatic})
into the corresponding formula for the LDOS in the quasistatic limit to second-order perturbation theory 
\begin{equation}
  \langle {D^\rE}^{(0)-(2)} \rangle_\rp \approx \frac{\omega}{\pi^2 c^2} \int_0^\infty \!\!\!\rd \kappa\, \re^{-2 \kappa z}
                                    \frac{\kappa^2}{2 k_0^2} \Im\bigl( \langle r_\rp^{(0) - (2))}\rangle_\rp \bigr).
\label{Eq:QuasiStaticLDOS}
\end{equation}
which follows from Eq.~(\ref{Eq:LDOS_sec_ord}) when assuming $\kappa \gg k_0$, 
we find
\begin{equation}
  \langle D^\rE (z) \rangle_\rp^{\rm PA} \approx  {D^\rE}^{(0)} \biggl(1 +6 \frac{\delta^2}{z^2} \biggr).
\label{Eq:PAresult}
\end{equation}
From this result the approximation for the radiative heat transfer in Eq.~(\ref{Eq:PArht}) follows easily.

Now, exactly the same result can be obtained with the so-called proximity approximation (PA), which holds 
in the quasistatic limit for $z \ll a$ and $z \gg \delta$  which was
for example used to determine the surface roughness contribution to the Casimir force~\cite{GenetEtAl2003,MaioNetoEtAl2005} and
has recently been employed to the near-field radiative heat transfer~\cite{PerssonEtAl2010}.
It amounts to replace the rough surface by a horizontal plane at a random height $z = S(\mathbf{x})$ followed by the
ensemble average. Hence, the LDOS reads: 
\begin{equation}
\begin{split}
  \langle D^{\rE} (z) \rangle_\rp &\approx \langle {D^\rE}^{(0)} (z - S(\mathbf{x})) \rangle_\rp \\
                                  &= {D^\rE}^{(0)}(z) + \frac{1}{2} \frac{\partial^2}{\partial z^2} {D^\rE}^{(0)} \delta^2
                                     +\mathcal{O}(4).
\end{split}
\end{equation} 
Utilizing the expression for the LDOS in the quasistatic limit~\cite{JoulainEtAl2003}
\begin{equation}
  {D^\rE}^{(0)} \approx \frac{1}{4} \frac{1}{(k_0 z)^3} \frac{\omega^2}{\pi^2 c^3} \frac{\epsilon''}{|\epsilon + 1|^2}
\label{Eq:QuasiStaticLimit}
\end{equation}
one retrieves the PA in Eq.~(\ref{Eq:PAresult}) when considering terms up to second order only.
Since the zeroth-order LDOS is proportional to $1/z^3$, it is clear that the contributions for planes at distances smaller
than $z$ will give a larger value than those at distances larger than $z$ so that after averaging the overall roughness 
correction is positive. Regarding only second-order terms it is also clear that this correction is proportional to $\delta^2$. 

\subsection{large distance regime ($\lambda_\rth \gg z \gg a$ and $a \ll \rd_\rs$)}

\begin{figure}[Hhbt]
  \epsfig{file=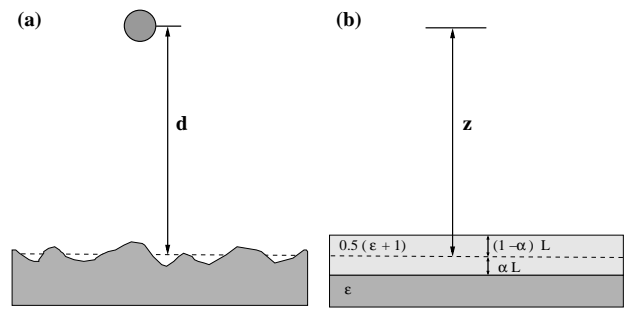, width=0.45\textwidth}
  \caption{\label{Fig:EffectiveLayer} Sketch of the physical situation encountered in the large distance regime for the heat transfer (a) and
                                      the replacement of the rough surface by an equivalent layer for surface roughness (b) to determine the LDOS in that regime.}
\end{figure}

When considering the evanescent contribution to the heat transfer in the large distance regime with $d \gg a$ one is facing a situation as sketched in 
Fig.~\ref{Fig:EffectiveLayer}~(a). At room temperature $\lambda_\rth \approx 10\,\mu{\rm m}$, so that
the evanescent regime is restricted to $d \ll \lambda_\rth = 10\,\mu{\rm m}$. Therefore,
this large distance regime is only applicable for surface profiles with a correlation 
length $a$ much smaller than $10\,\mu{\rm m}$ and a rms $\delta \ll a$. The same is true for the LDOS at $z\gg a$. 
On the other hand, for very small temperatures of about $5\,{\rm K}$, the thermal wavelength
is about $\lambda_\rth \approx 0.46\,{\rm mm}$. Hence, for low temperature experiments as the measurement of spin-flip lifetimes
in Ref.~\cite{Henkel2005,NoguesEtAl2009}, this large distance regime can be applied to a much wider range of surface roughness parameters, distances 
and nanoparticle radii.  
 
Considering the large distance limit $\kappa a \ll 1$ in Eq.~(\ref{Eq:Ms1}), we find for the lowest nonvanishing order  
\begin{align}
  M_{\rs,1} \approx - (k_0 \delta)^2 (\epsilon - 1)^2 \frac{\sqrt{\pi}}{2} \frac{1}{a} \frac{1}{\epsilon + 1}
\label{Eq:Mrs1largedist}
\end{align}
and with Eq.~(\ref{Eq:Mp1}) 
\begin{equation}
  M_{\rp,1} \approx \frac{\sqrt{\pi}}{2} (k_0 \delta)^2 \frac{(\epsilon - 1)^2}{\epsilon (\epsilon + 1)} \frac{1}{a}
                    \biggl(1 + \frac{\kappa^2}{k_0^2} \frac{2 \epsilon - 1}{\epsilon} \biggr).
\label{Eq:Mrp1largedist}
\end{equation}
Obviously, $M_{\rs/\rp,1}$ are proportional to $a^{-1}$. Therefore, for large distances for
which $|M_{\rs/\rp,1}| \gg |M_{\rs/\rp,0}|$ is fullfilled, the second-order correction to the LDOS
will be inversely proportional to the correlation length if $a \ll \rd_\rs$. The condition $|M_{\rs/\rp,1}| \gg |M_{\rs/\rp,0}|$ 
is fullfilled for the p-polarized modes in the here given limit $\kappa a \ll 1$, whereas for the
s-polarized modes, this is only true if $2 \gamma_\rt a (\epsilon + 1) \ll \sqrt{\pi}$. 
In addition, the approximate expressions in Eqs.~(\ref{Eq:Mrs1largedist}) and (\ref{Eq:Mrp1largedist}) have the 
denominator $\epsilon + 1$ indicating a strong contribution for surface resonances with $\epsilon' = -1$ 
and $\epsilon'' \approx 0$. Therefore, for frequencies near the surface resonance frequency and for distances
$z \gg a$ the proper self energy $M_{\rs/\rp}$ in the expressions for the reflection coefficient in Eq.~(\ref{Eq:reflspec}) 
can be replaced by Eqs.~(\ref{Eq:Mrs1largedist}) and (\ref{Eq:Mrp1largedist})
yielding  the large distance approximation (LDA).

As far as one considers correlation lengths $a$ much smaller than the wavelength inside the medium $\lambda/|\sqrt{\epsilon}|$,
one can apply the concept of homogenization~\cite{BrundrettEtAl1994} 
by replacing the surface roughness by an equivalent surface layer with an effective permittivity. In the near-field
regime, this condition is fullfilled if $\kappa a \ll 1$ or $a \ll z$, resp. 
Following the approach of Rahman and Maradudin~\cite{RahmanMaradudin1980a,RahmanMaradudin1980b}
we replace 
the rough surface in the limit of large distances with $z \gg a$ and for $a \ll \rd_\rs$  by a thin 
equivalent surface layer as depicted in Fig.~\ref{Fig:EffectiveLayer} ranging 
from $z = - \alpha L$ to $(1 - \alpha) L$ with $\alpha \in [0,1]$. The permittivity $\epsilon_\rL$ of that layer is 
considered to be $0.5(\epsilon + 1)$, i.e., the average of the permittivity of vacuum and the dielectric medium. 
The reflection coefficient for such a layered geometry with $\epsilon_1 = \epsilon$, $\epsilon_2 = \epsilon_\rL$, and
$\epsilon_3 = 1$ is in the quasistatic limit $\kappa \gg k_0$
\begin{equation}
  r_{\rs/\rp}^\rL \approx \re^{2 \kappa (1 - \alpha) L} \frac{r^{32}_{\rs/\rp} + r^{21}_{\rs/\rp} \re^{- 2 \kappa L}}{1 + r^{32}_{\rs/\rp} r^{21}_{\rs/\rp} \re^{2 \kappa L}}.
\end{equation}

Considering first p-polarised modes we insert the electrostatic expressions for the Fresnel coefficients
\begin{equation}
  r^{32}_\rp \approx \frac{\epsilon_\rL - 1}{\epsilon_\rL + 1} \qquad\text{and}\qquad
  r^{21}_\rp \approx \frac{\epsilon - \epsilon_\rL}{\epsilon + \epsilon_\rL}
\end{equation}
and expand the reflection coefficient for the layered system with respect to the thickness $L$, which is
thought to be small, yielding
\begin{equation}
   r_{\rp}^\rL \approx \frac{\epsilon - 1}{\epsilon + 1} 
                           - \kappa L\frac{(\epsilon - 1)}{(\epsilon + 1)^3}\bigl[ - \alpha (1 + \epsilon)^2 + \epsilon^2 + \epsilon + 1 \bigr].
\end{equation}

Now, we want to compare $\Delta r_\rp^\rL  = r_\rp - r_{\rp}^\rL$ with the result of the LDA. 
Inserting Eq.~(\ref{Eq:Mrp1largedist}) into the expression for the mean reflection coefficients
in Eq.~(\ref{Eq:reflspec}) gives for the quasistatic limit ($\kappa \gg k_0$)
\begin{equation}
  \Delta \langle r_{\rp}^{(0)-(2)} \rangle_\rp = \langle r_{\rp}^{(0)-(2)} \rangle_\rp  - r_{\rp} 
                      \approx \frac{\sqrt{\pi}}{2} \frac{ \kappa\delta^2}{a} \frac{(\epsilon - 1)^2}{(\epsilon + 1)^3} (2 \epsilon - 1).
\label{Eq:DeltaRLDA}
\end{equation}
In order to get a relation between $L$ and $\alpha$, we compare the leading term in $\epsilon$ for $\delta r_\rp^\rL$ of 
the layered geometry with 
the surface roughness correction. This gives
\begin{equation}
  (1 - \alpha) L = \frac{\sqrt{\pi} \delta^2}{a}.
\label{Eq:ELp}
\end{equation} 
Obviously, the choice of the parameters $\alpha$ and $L$ is ambiguous. This ambiguity is resolved in Ref.~\cite{RahmanMaradudin1980a} 
when further considering the transmitted field component for the equivalent layer and the corresponding perturbative result. 
From this, Rahman and
Maradudin find $L = 3 \sqrt{\pi} \delta^2/a$.
Using this result in Eq.~(\ref{Eq:ELp}) gives $\alpha = 2/3$ as was also found by Rahman and Maradudin in Ref.~\cite{RahmanMaradudin1980a} 
Hence, the surface roughness can be mimicked
by a small layer with a thickness of order $\delta^2/a$ which is slightly 
shifted into the vacuum region with respect to $z = 0$, so that the effective distance from the surface becomes $z - L/3$. It follows, that for distances $z \gg a$ the
evanescent part of the LDOS and of the heat transfer, which have a monotonous decay with $z$, will be slightly bigger than that of a flat surface and proportional to $a^{-1}$.

The same considerations can be made for the s-polarized modes yielding the results
\begin{equation}
  \Delta r_\rs^\rL \approx (k_0 \delta)^2 \frac{1}{\kappa a} \frac{3}{2} \sqrt{\pi}\biggl[ \frac{1}{2}(1 - 5 \epsilon) -\alpha (1 - \epsilon) \biggr]
\end{equation}
in the quasistatic regime, which can be compared to the corresponding correction to the
mean reflection coefficient (by inserting Eq.~(\ref{Eq:Mrs1largedist}) into Eq.~(\ref{Eq:reflspec}))
\begin{equation}
  \Delta \langle r_{\rs}^{(0)-(2)} \rangle_\rp  \approx - (k_0 \delta)^2 \frac{\sqrt{\pi}}{2} \frac{1}{\kappa a} \frac{1}{4} \frac{(\epsilon - 1)^2}{(\epsilon + 1)} .
\end{equation}
Comparing the leading order term of $\delta r_\rs^\rL$ and $\delta \langle r_{\rs}^{(0)-(2)} \rangle_\rp$ in $\epsilon$ gives 
\begin{equation}
  (5 - 2 \alpha) L = \frac{\sqrt{\pi} \delta^2}{2 a}.
\end{equation}
Again, the choice of $\alpha$ and $L$ is ambiguous. Using again the result for $L$ from 
Ref.~\cite{RahmanMaradudin1980a} gives $\alpha = 29/12$.  
Obviously, $\alpha > 1$ so that it is not possible to describe the effect of surface roughness for s-polarized modes within this
model when choosing the same thickness $L$ as for p-polarized modes. In order to get the correct $L$, it is necessary to consider 
the transmitted field contribution. Therefore, here it can only be concluded that for the s-polarized part we can also use the equivalent
layer model with $L$ of the order $\delta^2/a$.

As it might be shown in the next section, in the large distance regime, the roughness plays a significant role only at the surface polariton resonance, i.e.,
for $\omega_\rs$ such that $\Re(\epsilon(\omega_\rs)) + 1 \approx 0$ and $\Im(\epsilon(\omega_\rs)) \approx 0$. This is clear when noting the resonant 
denominator $1/(\epsilon + 1)^3$ in Eq.~(\ref{Eq:DeltaRLDA}). Hence, we expect a significant modification of the LDOS and therefore also of the lifetime of an 
emitter whose frequency is close to $\omega_\rs$. 
On the other hand, for the near-field heat transfer one has to sum up all contributions over the spectrum close to the thermal frequency so that
one can in general expect that the correction is in this case small. 

Finally, we note that, although the physics is different, the effect of roughness on heat transfer is the same for evanescent and propagating waves. The
roughness can be modeled by an effective layer with intermediate optical properties. In both cases, it results in a larger transmission.

%
%

\section{Numerical results for the LDOS}

\subsection{Propagating modes}
\label{Sec:ReflProp}

\begin{figure}[Hhbt]
  \epsfig{file=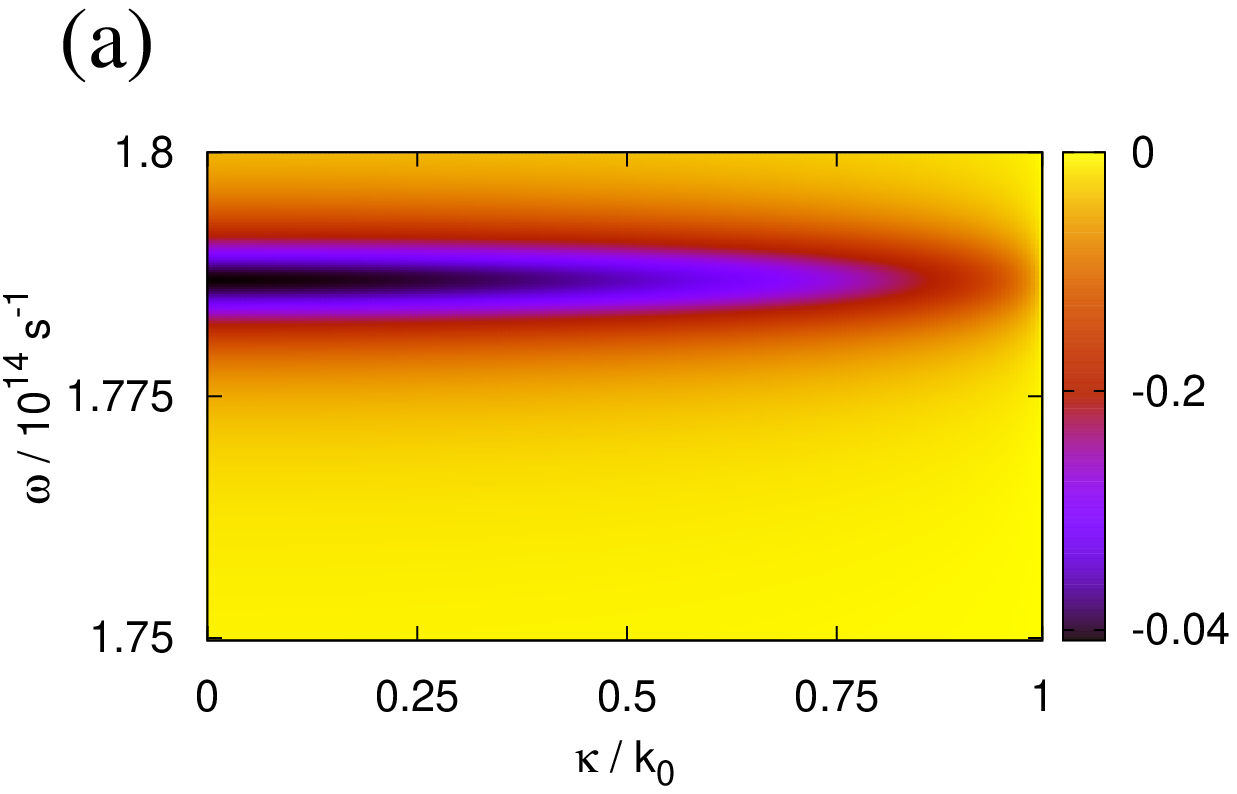, width=0.35\textwidth}
  \epsfig{file=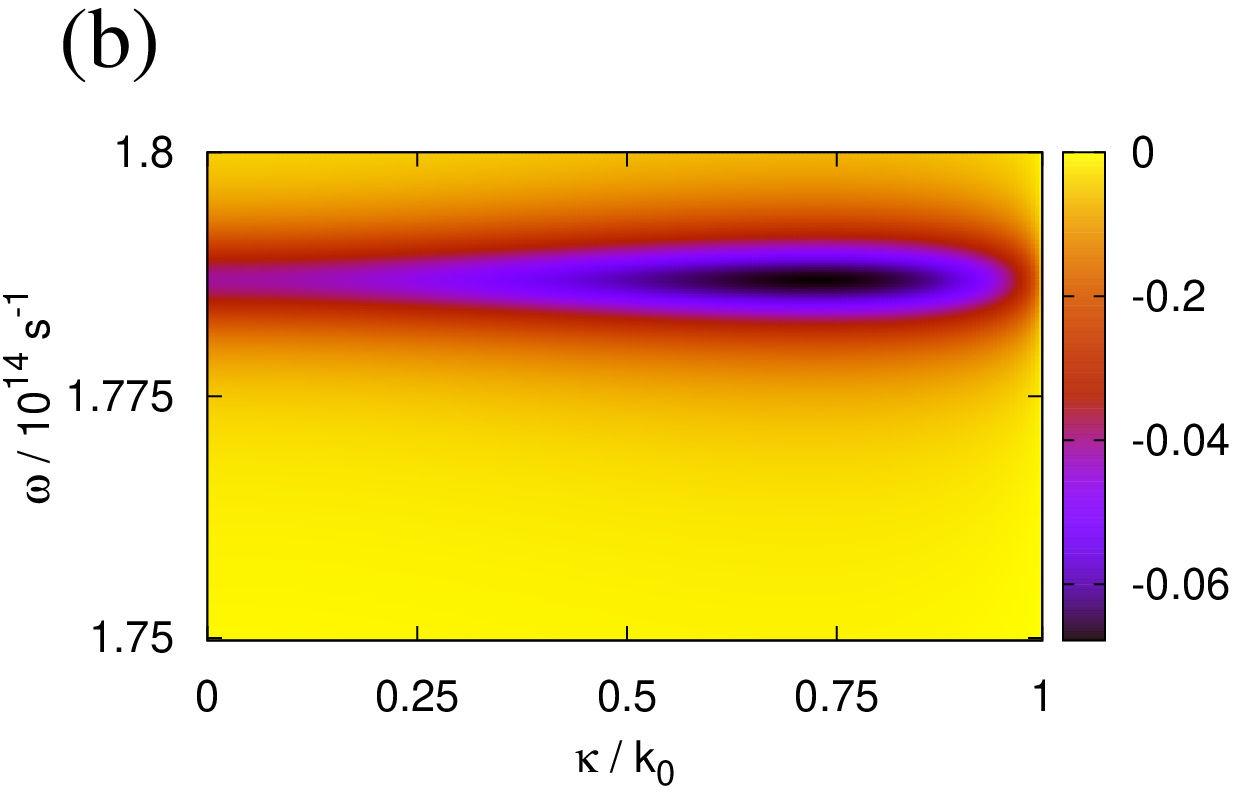, width=0.35\textwidth}
  \caption{\label{Fig:Reflspec} (Color online)
           Plot of $\Delta R_{\rs/\rp}$ as defined in 
           Eq.~(\ref{Eq:DeltaRefl}) for
           (a) s- and (b) p-polarized modes using $\delta = 5\,{\rm nm}$ and $a = 200\,{\rm nm}$.}
\end{figure}

First, we discuss some numerical results for the propagating modes
using the material parameters for SiC from Ref.~\cite{ShchegrovEtAl2000}. In
Fig.~\ref{Fig:Reflspec}, we show a plot of the deviation of reflectivity from Eq.~(\ref{Eq:reflspec}) 
defined as
\begin{equation}
  \Delta R_{\rs/\rp} =  100 \, \frac{|\langle r_{\rs/\rp}^{(0)-(2)}\rangle_\rp|^2 - |r_{\rs/\rp}|^2}{|r_{\rs/\rp}|^2}
\label{Eq:DeltaRefl}
\end{equation}
for s- and p-polarization for $\kappa \leq k_0$ choosing a rms of $\delta = 5\,{\rm nm}$
and a correlation length of $a = 200\,{\rm nm}$. Furthermore, the plot is restricted to 
frequencies around the surface phononon polariton resonance $\omega_\rs = 1.787\cdot 10^{14}\,{\rm s}^{-1}$ within the 
reststrahlenband, i.e., $\omega_\rt < \omega < \omega_\rl$, where $\omega_\rl = 1.827\cdot 10^{14}\,{\rm s}^{-1}$ and
$\omega_\rt = 1.495\cdot 10^{14}\,{\rm s}^{-1}$ are the frequencies of the LO and TO phonons in SiC, resp. It can be seen that
the correction is small and negative in this frequency range. For frequencies around the
surface phonon frequency one finds a relatively large negative correction (but still smaller
than one percent) for both polarisations. Hence, the roughness correction to the LDOS for propagating modes
will be very small as well.     

The surface roughness allows for coupling of incident propagating waves 
with surface polaritons~\cite{MaradudinMills1975,MarvinEtAl1975,Agarwal1977}. Hence, for $(\omega,\kappa)$ for which the conditions for coupling 
of surface polaritons with propagating modes are met, the reflectivity will decrease, i.e.,
$\Delta R_{\rp}$ is negative, since only a small fraction of the excited surface mode
will be reradiated.  
This surface wave mediated decrease of the reflectivity can be seen for both s- and p-polarization. 

\begin{figure}[Hhbt]
  \epsfig{file=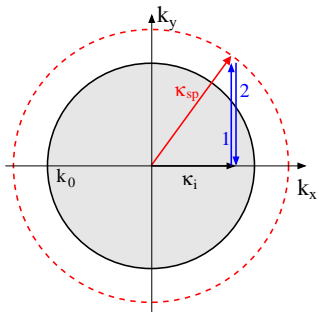, width=0.23\textwidth}
  \caption{\label{Fig:Phasematch} (Color online)
           Illustration of the excitation of surface polaritons by s-polarized light in the $\boldsymbol{\kappa}$-plane
           (dashed line = dispersion relation of the surface phonon polariton, gray circle = propagating waves with $\kappa < k_0$): 
           An s-polarized wave with $\boldsymbol{\kappa}_\ri = (k_\ri,0)$ and an electric field 
           in $y$ direction.
           Due to a first scattering process labeled as $1$ with the rough surface the surface power spectrum 
           provides the necessary extra momentum $(\boldsymbol{\kappa}_\ri - \boldsymbol{\kappa}_{\rm sp})$
           to match the phase with the surface polariton with wave vector $\boldsymbol{\kappa}_{\rm sp}$.  The incoming s-polarized wave has an electric field component in the direction of $\boldsymbol{\kappa}_{\rm sp}$ so that it can excite
           the surface polariton. Due to the second scattering process $2$ the surface power spectrum again provides     
           the necessary extra momentum $-(\boldsymbol{\kappa}_\ri - \boldsymbol{\kappa}_{\rm sp})$ resulting in 
           a scattered s-polarized wave with $\boldsymbol{\kappa} = \boldsymbol{\kappa}_\ri$.
           }
\end{figure}

The possibility of exciting surface polaritons with s-polarized light needs some explanation, 
since after taking the ensemble average, the rough surface has the same symmetries as a flat surface for which surface
polaritons can be excited with p-polarised light only. On the other hand, for surfaces with a grating 
which breaks the translational and rotational symmetry, it is known that~\cite{ElstonEtAl1991,MarquierEtAl2008} s-polarized waves 
can excite surface polaritons.
In this case, the grating provides the extra momentum for matching the phase of the incoming light with that 
of the surface polariton. Additionally, the incoming wave must have an electric field vector component parallel
to the surface phonon polariton wave vector. When these two conditions
are met~\cite{ElstonEtAl1991,MarquierEtAl2008}, an s-polarized wave can excite surface polaritons.
Now, for a rough surface the translational and rotational symmetry are also broken 
so that an s-polarized wave should be able to excite 
surface polaritons. An example of such a scattering process is illustrated and discussed in Fig.~\ref{Fig:Phasematch}.
 
\subsection{Evanescent modes}
\label{Sec:ReflEv}

\begin{figure}[Hhbt]
  \epsfig{file=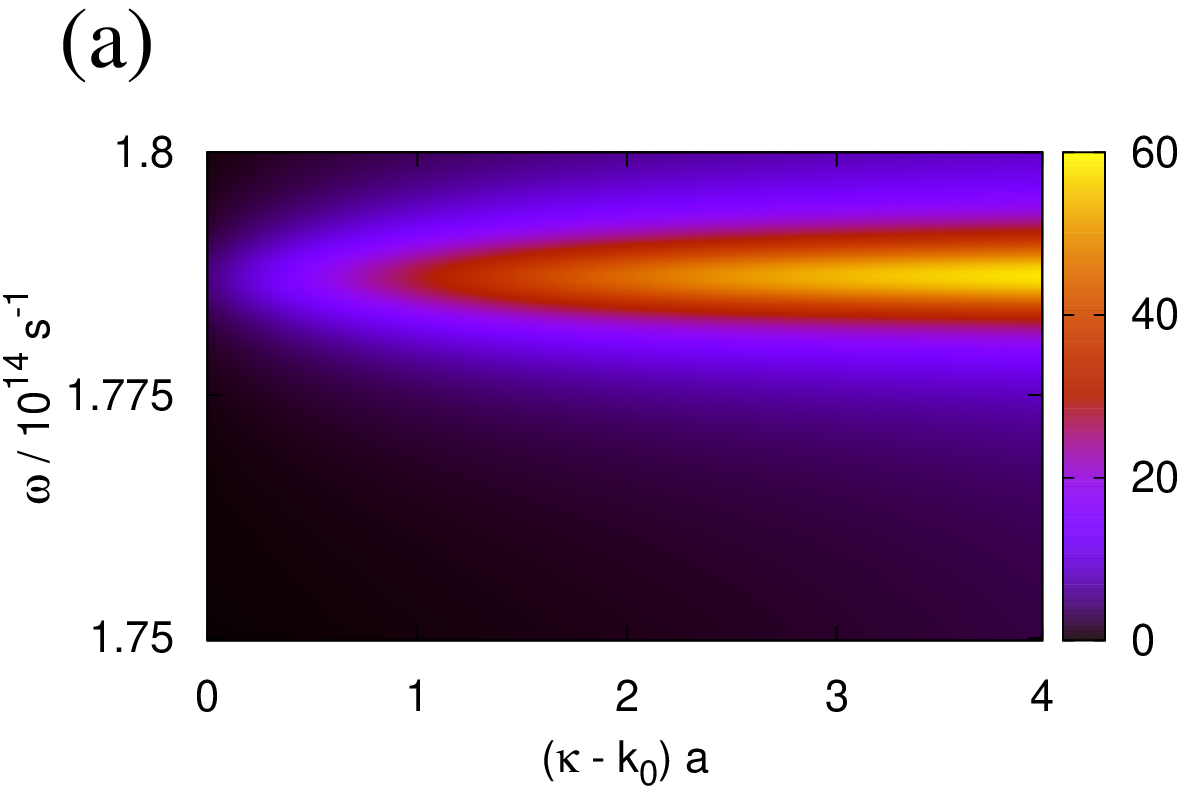, width=0.35\textwidth}
  \epsfig{file=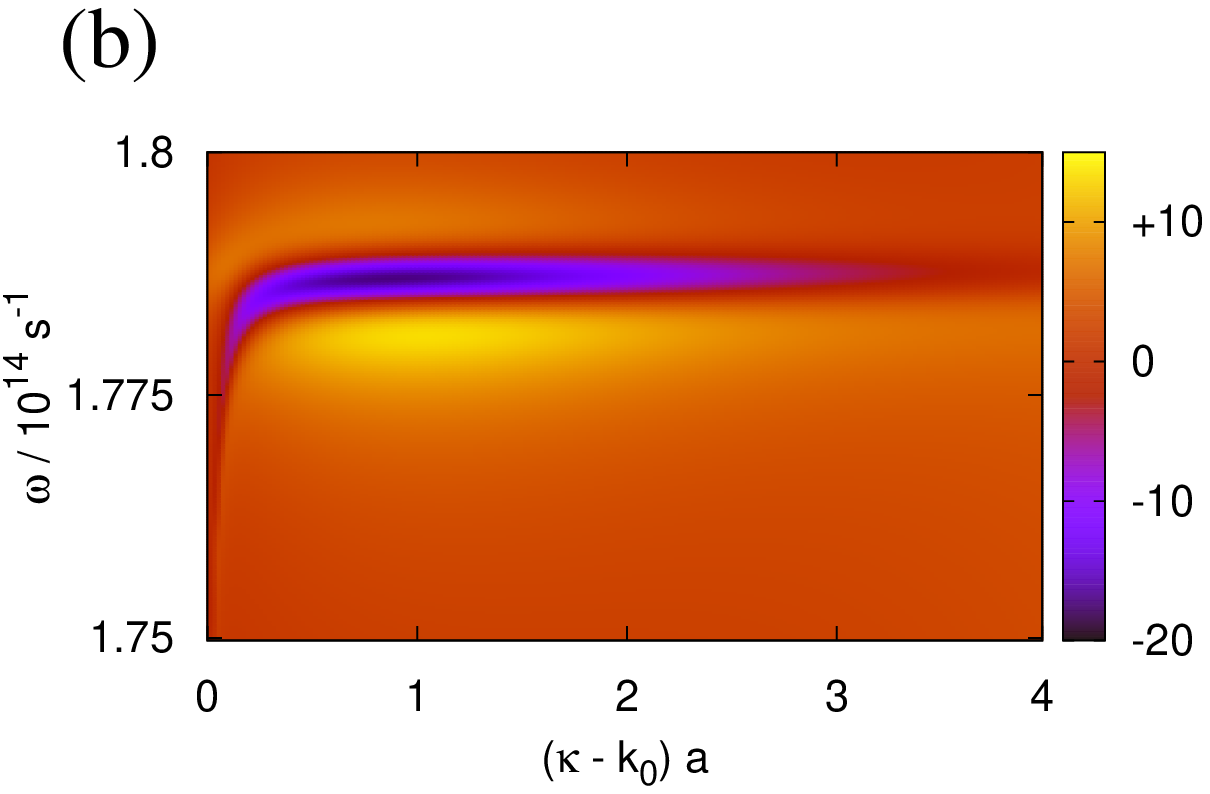, width=0.35\textwidth}
  \caption{\label{Fig:ImRefl} (Color online)
           Plot of $\Im( r_{\rs/\rp})$ as defined in 
           Eq.~(\ref{Eq:DeltaImRefl}) for  
           (a) s- and (b) p-polarized modes using $\delta = 5\,{\rm nm}$ and $a = 200\,{\rm nm}$.}
\end{figure}

Now, we turn to the evanescent modes. According to Eq.~(\ref{Eq:LDOS_sec_ord}), we are interested in $\Im(r)$. In order to discuss the change in $\Im(r)$ we
define 
\begin{equation}
  \Delta \Im(r_{\rs/\rp}) =  100\, \frac{\Im(\langle r_{\rs/\rp}^{(0)-(2)} \rangle_\rp) - \Im(r_{\rs/\rp}) }{\Im(r_{\rs/\rp})}.
\label{Eq:DeltaImRefl}
\end{equation}
In Fig.~\ref{Fig:ImRefl} we show a plot of this quantity for s- and p-polarized modes using again the
parameters $\delta = 5\,{\rm nm}$ and $a = 200\,{\rm nm}$. We show here only the results
for $\kappa a \leq 4$, i.e., in that region where the mean reflection coefficient is 
dominated by $M_{\rs/\rp,1}$. For s-polarized modes one can see that
$\Im(r_\rs)$ is increased up to $60$ percent for values of $(\omega,\kappa)$  
coinciding with the dispersion relation of surface phonons. The underlying mechanism is the same as for the propagating modes depicted
in Fig.~\ref{Fig:Phasematch}, but with the difference that $\boldsymbol{\kappa}_\ri$ is greater than $k_0$.   

\begin{figure}[Hhbt]
  \epsfig{file=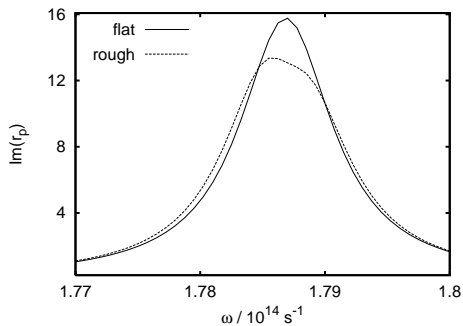, width=0.35\textwidth}
  \caption{\label{Fig:ImReflb}
           Plot of $\Im(r_\rp)$ and $\Im(\langle r_\rp^{(0)-(2)} \rangle_\rp)$ for 
           $\kappa a = 1$ using $\delta = 5\,{\rm nm}$ and $a = 200\,{\rm nm}$.} 
\end{figure}

On the other hand, for p-polarized modes $\Im(r_\rp)$ is decreased by about $20$ percent 
for values of $(\omega,\kappa)$ coinciding with the dispersion relation of the surface phonons. 
Around $\kappa a \approx 1$ one also finds a large increase of about $15$ percent for frequencies
slighty below and above the surface phonon frequency. This can be easily understood by looking 
at Fig.~\ref{Fig:ImReflb} showing a plot of $\Im(r_\rp)$ and $\Im(\langle r_\rp^{(0)-(2)} \rangle)$ for
$\kappa a = 1$ and for frequencies near the surface phonon frequency. As can be observed, 
the dispersion relation is broadened due to roughness induced scattering
of the surface phonons into other surface phonon states. For a slightly rougher surface with
$\delta = 10\,{\rm nm}$ the broadening becomes a splitting of the surface polariton
dispersion~\cite{RahmanMaradudin1980b,MoragaLabbe1989}, which can be easily understood from the fact that
the rough surface acts as a thin layer as discussed above.  It follows, that the quantity $\Delta \Im(r_{\rs/\rp})$ 
has negative values for frequencies around the surface phonon frequency and positive values slightly 
below and above that surface phonon frequency as can be seen in Fig.~\ref{Fig:ImRefl}. We remark that
for $\delta = 10\,{\rm nm}$ the direct perturbation theory, while qualitatively correct, starts to give quantitatively wrong results for frequencies near the surface resonance frequency~\cite{FariasMaradudin1983}. 

\subsection{Spectrum of the LDOS}

From the above discussion it is clear that the roughness correction of the LDOS defined as
\begin{equation}
  \Delta D^\rE =  100 \, \frac{\langle {D^\rE}^{(0)-(2)}\rangle_\rP - {D^\rE}^{(0)} }{{D^\rE}^{(0)}}
\label{Eq:DeltaDE}
\end{equation}
will have positive and negative contributions for different frequencies and distances.
To make this point clear, we show some plots of $\Delta D^\rE$ and $D^\rE =\langle {D^\rE}^{(0)-(2)}\rangle_\rP$ for frequencies
ranging from $10^{13}\,{\rm s}^{-1}$ to $2.5\cdot10^{14}\,{\rm s}^{-1}$. 

\begin{figure}[Hhbt]
  \epsfig{file=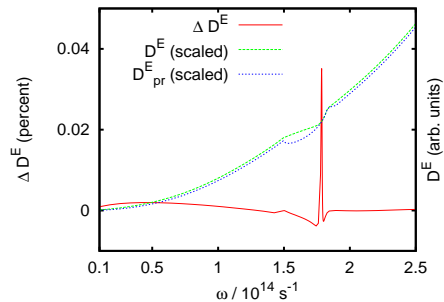, width=0.35\textwidth}
  \caption{\label{Fig:DEspecfar} (Color online)
           Plot of $D^\rE$ and $\Delta D^\rE$ as defined in 
           Eq.~(\ref{Eq:LDOS_sec_ord}) and Eq.~(\ref{Eq:DeltaDE}) for  
           propagating and evanescent modes (dashed line) and for propagating modes only (dotted line) using  
           $\delta = 5\,{\rm nm}$ and $a = 200\,{\rm nm}$ and
           distance $z = 5\,\mu {\rm m}$.}
\end{figure}

In Fig.~\ref{Fig:DEspecfar} we plotted $D^\rE$ and $\Delta D^\rE$ for propagating and evanescent modes 
and for propagating modes only for a distance of $z = 5\,\mu{\rm m}$. It is seen that $D^\rE$ is
dominated by the propagating modes and shows a small dip in the reststrahlenband of SiC, where the contribution
of the evanescent modes is already relatively large. The surface roughness correction is in this case
extremely small and originates in equal parts from the correction to the evanescent and propagating modes. 

\begin{figure}[Hhbt]
  \epsfig{file=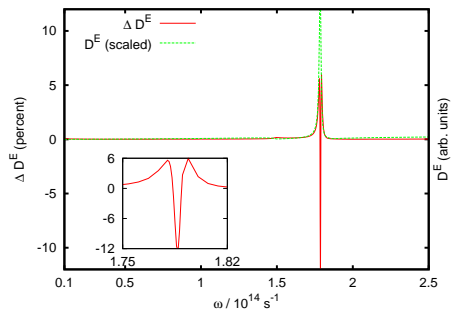, width=0.35\textwidth}
  \caption{\label{Fig:DEspecnear} (Color online)
           Plot of $D^\rE$ and $\Delta D^\rE$ as defined in 
           Eq.~(\ref{Eq:DeltaDE}) for $z = 500\, {\rm nm}$ using the same roughness parameters as in Fig.~\ref{Fig:DEspecfar}.}
\end{figure}

For very small distances the LDOS is solely dominated by the evanescent contribution so that in this case 
the spectrum is quite different from that of the propagating part~\cite{JoulainEtAl2003,ShchegrovEtAl2000}. 
In Fig.~\ref{Fig:DEspecnear} we show a plot
of $D^\rE$ for $z = 500\,{\rm nm}$, i.e., in the evanescent regime, leaving the other parameters unchanged. As can be seen, 
the spectrum of $D^\rE$ has a resonance due to surface phonons. The curve of $\Delta D^\rE$
shows a negative deviation of about 12 percent at the surface phonon resonance and a
positive deviation of about 6 percent slightly below and above that resonance as can be seen in the inset. This 
behaviour is due to the scattering of surface phonons discussed in the previous section resulting in 
a broadened surface phonon dispersion.  

\subsection{Distance dependence of the LDOS}

\begin{figure}[Hhbt]
  \epsfig{file=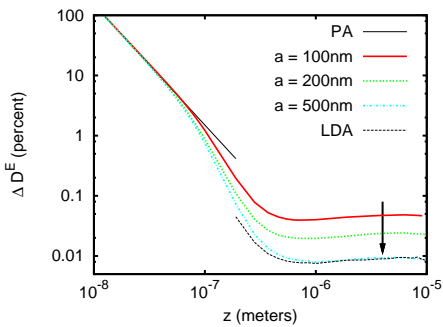, width=0.35\textwidth}
  \caption{\label{Fig:DeltaDE1e14} (Color online)
           Plot of $\Delta {D^\rE}$ over the distance for SiC using the frequency  
           $\omega = 10^{14}\,{\rm s}^{-1}$ for a rough surface with $\delta = 5\,{\rm nm}$ 
           for correlation length $a = 100\,{\rm nm}$, $200\,{\rm nm}$, and $ 500\,{\rm nm}$.
           The thin solid line represents the PA result, whereas the thin dash-dotted line is the
           large distance approximation (LDA).}
\end{figure}

Let us now focus on the distance dependence of the LDOS. For rough surfaces it can be expected 
that the roughness correction of the LDOS $\Delta D^\rE$ defined in Eq.~(\ref{Eq:DeltaDE}) converges for small distances, i.e., for $z \ll a$,
to the PA result. On the other hand, for large distances, i.e., for $z \gg a$, this correction should be inversely proportional to 
the correlation length and can be described quantitatively utilizing the LDA of the proper self energy in
Eqs.~(\ref{Eq:Mrs1largedist}) and (\ref{Eq:Mrp1largedist}). In Fig.~\ref{Fig:DeltaDE1e14} we show a plot of $\Delta D^\rE$ for SiC 
at the frequency $\omega = 10^{14}\,{\rm s}^{-1}$
using the roughness parameters $\delta = 5\,{\rm nm}$ and $a = 100\,{\rm nm}$, $200\,{\rm nm}$ and $ 500\,{\rm nm}$. 
It can be seen that the LDOS converges to the two limits for small and large distances giving a surface roughness correction
on the order of some percent only for distances smaller than $100\,{\rm nm}$. The arrow indicates that the surface
roughness correction decreases when increasing the correlation length $a$. 

\begin{figure}[Hhbt]
  \epsfig{file=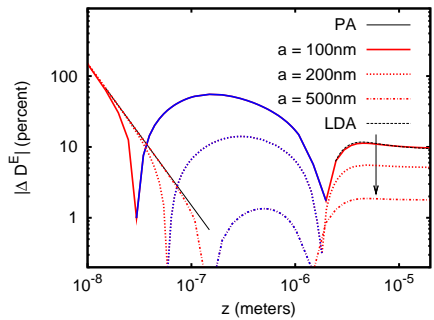, width=0.35\textwidth}
  \caption{\label{Fig:DeltaDE1787e11} (Color online)
           As Fig.~\ref{Fig:DeltaDE1e14} but for $\omega = 1.787 \cdot 10^{14}\,{\rm s}^{-1}$. The red part of the curves
           indicates positive values and the blue one negative values.}
\end{figure}

In Fig.~\ref{Fig:DeltaDE1787e11} we show a similar plot when choosing the surface phonon 
frequency $\omega = \omega_\rs = 1.787\cdot10^{14}\,{\rm s}^{-1}$ and plotting the modulus of $\Delta D^\rE$. 
As before, it can be observed that $\Delta D^\rE$ converges to the PA and the LDA 
for $z \ll a$ or $z \gg a$, resp. Apart from that for distances $z \approx a$ the surface roughness
correction to the LDOS $D^\rE$ becomes negative, a feature not seen for $\omega = 10^{14}\,{\rm s}^{-1}$.    
This behaviour can be understood from the above given discussion: the dispersion relation of the surface
phonons is broadened due to roughness induced scattering yielding a negative $\Delta \Im (r_\rp)$ for frequencies
around the surface phonon frequency and for wavevectors around $a^{-1}$, because in that region the roughness induced
scattering of surface phonons is strong. Since for a given distance $z$ the main contributions to $D^\rE$ stem from 
wavevectors $\kappa \approx z^{-1}$, this roughness induced broadening becomes important for distances $z \approx a$
giving a smaller LDOS than for a flat surface so that $\Delta D^\rE$ is negative. As can be seen in Fig.~\ref{Fig:DeltaDE1787e11} 
the correction varies from about $+ 10$ percent to about $-50$ percent in the distance regime $z > 100\,{\rm nm}$ for 
$a = 100\,{\rm nm}$.
 
\begin{figure}[Hhbt]
  \epsfig{file=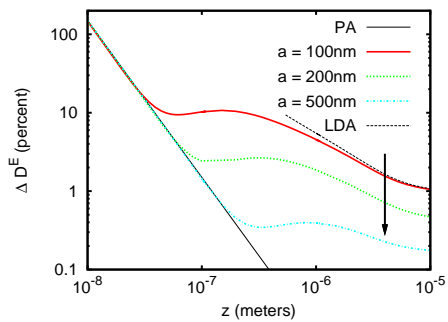, width=0.35\textwidth}
  \caption{\label{Fig:DeltaDE18e13} (Color online)
           As Fig.~\ref{Fig:DeltaDE1e14} but with $\omega = 1.8 \cdot 10^{14}\,{\rm s}^{-1}$.}
\end{figure}

Now, as can for example be seen for $\Delta \Im(r_p)$ in Fig.~\ref{Fig:ImRefl}~(b) and for the spectrum of $\Delta D^\rE$ in 
Fig.~\ref{Fig:DEspecnear} for frequencies slightly below or above the surface resonance the LDOS increases due to the
roughness induced scattering of surface phonons. Hence, relatively large positive surface roughness correction can be 
expected for such frequencies. To illustrate this statement we also plot $\Delta D^\rE$ for $\omega = 1.8\cdot10^{14}\,{\rm s}^{-1}$
in Fig.~\ref{Fig:DeltaDE18e13}.

It is apparent from the above given examples that the distance dependence of the surface roughness correction to the LDOS 
is not only sensitive to the surface roughness parameters themself, but also to the chosen frequency. Additionally, the
surface roughness correction $\Delta D^\rE$ is in general nonmonotonous and especially large for frequencies around the surface phonon resonance. 

%
%

\section{Summary}

In this work, we studied the near-field radiative heat transfer between a nanoparticle and
a rough surface utilizing direct perturbation theory up to the lowest nonvanishing order
in the surface profile. Employing the material properties of SiC we have shown
that the distance dependence of the roughness correction to the heat flux is nonmonotonous and can be qualitatively 
understood from the roughness correction to the LDOS. 

We have derived an approximation in the small distance regime, i.e., for distances $d$ much smaller than the correlation length $a$,
and have shown that it exactly coincides with the results of the proximity approximation.
Hence, the roughness correction is well described by the PA for $d \ll a$. 
Therefrom, one can conclude that the PA might also be helpful in other geometries as
used in recent experimental setups~\cite{HuEtAl2008,NarayaEtAl2008,RousseauEtAl2009,ShenEtAl2008} to estimate the impact of surface roughness to the 
near-field radiative heat transfer at small distances. Since, the numerical results give larger values for the heat transfer than predicted by the PA
for most distances $d$ , one can expect to get an estimate of the lower limit of the surface roughness correction when using the PA.

In the large distance limit $d \gg a$ we have derived a simple approximation for the corresponding expressions
of the LDOS and the heat flux, which simplifies the numerical calculations in this limit tremendously. 
We have shown that in this regime the rough surface can be replaced by an equivalent surface layer of thickness $\delta^2/a$
for correlation lengths $a$ smaller than the skindepth making contact with the results 
of Refs.~\cite{RahmanMaradudin1980a,RahmanMaradudin1980a} The corrections to the heat
transfer are therefore relatively small in that regime when considering roughness with $a \gg \delta$.  

In the intermediate regime we could show that the LDOS and heat flux for a rough material can be 
smaller than that of a material with a flat surface due to the roughness induced scattering of surface phonon polaritons.
Furthermore, we have shown that due to surface roughness the LDOS and therefore the heat flux has an s-polarized surface polariton
contribution. The mechanisms behind these two unexpected results has been discussed. 

Finally, we want to emphasize that the results for the LDOS presented in that work have a much larger range of applicability, since
the LDOS, for instance, also determines the lifetime of atoms and molecules near a surface.   

%
%

\begin{acknowledgments}
S.-A.\ B. gratefully acknowledges support from the Deutsche Akademie der Naturforscher Leopoldina
(Grant No.\ LPDS 2009-7).
\end{acknowledgments}

%
%

\appendix

\section{Perturbation result for Green's function}

With the procedure in Ref.~\cite{HenkelSandoghdar1998} and the perturbation theory in Ref.~\cite{Greffet1988} 
it is possible to determine the Green's dyadic 
for each order with respect to the surface profile. The zeroth- and first order expressions 
can be found in Ref.~\cite{HenkelSandoghdar1998}. For the second-order Green's dyadic we find for $z < z'$
\begin{equation}
  {\mathds{G}^\rE}^{(2)} (\mathbf{r,r'}) = \frac{1}{2}\int\!\! \frac{\rd^2 \kappa}{(2 \pi)^2}\, \re^{\ri(\boldsymbol{\kappa}\cdot\mathbf{x} + \gamma_\rr z)} {\mathds{G}^\rE}^{(2)}(\boldsymbol{\kappa})
\label{Eq:GreensfunctionAppendix}
\end{equation}
with
\begin{equation}
  \begin{split}
    {\mathds{G}^\rE}^{(2)}(\boldsymbol{\kappa}) &= k_0^2 \frac{1 - \epsilon}{2 \gamma_\rr} \biggl[\int\!\!\frac{\rd^2 \kappa'}{(2 \pi)^2}\,
                                                    \tilde{S}^{(2)}(\boldsymbol{\kappa}' - \boldsymbol{\kappa}) \\
                                                &\qquad\times    \mathcal{T}({k_\rr^+},{k_\rr^+}) \mathds{N}(k_\rr^+,{k_\rt^-}')
                                                    \mathcal{T}({k_\rt^-}',{k_\rr^-}') f(\boldsymbol{\kappa}',z')\\
                           &+  \int\!\!\frac{\rd^2 \kappa'}{(2 \pi)^2} 
                                      \int\!\!\frac{\rd^2 \kappa''}{(2 \pi)^2} \,
                                      \tilde{S}^{(1)}(\boldsymbol{\kappa}' - \boldsymbol{\kappa})  \tilde{S}^{(1)}(\boldsymbol{\kappa}'' - \boldsymbol{\kappa}') \\
                           &\qquad\times  2 (\gamma_\rt' - \gamma_\rr') \mathcal{T}({k_\rr^+},{k_\rr^+}) \mathds{L}(k_\rr^+,{k_\rt^-}',{k_\rt^-}'') \\
                           &\qquad\times  \mathcal{T}({k_\rt^-}'',{k_\rr^-}'') f(\boldsymbol{\kappa}'',z')  \biggr] 
\end{split}
\end{equation}
where
\begin{align}
  f(\boldsymbol{\kappa},z')        &= \frac{\ri}{2 \gamma_\rr} \re^{- \ri \boldsymbol{\kappa}\cdot\mathbf{x}' + \ri \gamma_\rr z'}, \\
  \tilde{S}^{(n)}(\boldsymbol{\kappa} - \boldsymbol{\kappa}') &= \int\!\!\rd^2 x \, \re^{\ri \mathbf{x}\cdot(\boldsymbol{\kappa} - \boldsymbol{\kappa}')} \bigl(S(\mathbf{x})\bigr)^n 
\end{align} 
for $n = 1,2$.
Here all matrices $\mathds{A}$ are defined by
\begin{equation}
  \mathds{A}_{i,j}(k_\rr^\pm,{k_\rt^\pm}') = A_{ij} \hat{\mathbf{a}}_i(k_\rr^\pm) \otimes \hat{\mathbf{a}}_j({k_\rt^\pm}') \label{Eq:DefR}, 
\end{equation}
where $i,j = (\rs,\rp)$ and 
\begin{equation}
  \hat{\mathbf{a}}_\rs(k^\pm) = \hat{\mathbf{z}} \times \hat{\boldsymbol{\kappa}} \quad\text{and}\quad
  \hat{\mathbf{a}}_\rp(k^\pm) = \hat{\mathbf{a}}_\rs \times \frac{\mathbf{k}^\pm}{|\mathbf{k}^\pm|}, 
\end{equation}
are the normalized and orthogonal polarization vectors;
$\hat{\mathbf{z}}$ is the unit vector in z-direction, $\hat{\boldsymbol{\kappa}} = \boldsymbol{\kappa}/\kappa$ and 
$\mathbf{k}^\pm = (\boldsymbol{\kappa},\pm \gamma)^t$. 

The matrix $\mathcal{T}$ is given by the diagonal matrix
\begin{equation}
  \mathcal{T} = \begin{pmatrix} \frac{2 \gamma_\rr}{\gamma_\rr + \gamma_\rt} & 0 \\ 
                               0 & \frac{2 \sqrt{\epsilon} \gamma_\rr}{\epsilon \gamma_\rr + \gamma_\rt} \end{pmatrix} 
             = \begin{pmatrix} t_\rs & 0 \\ 
                               0 & t_\rp \end{pmatrix} 
\end{equation}
defining the amplitude transmission coefficients $t_\rs$ and $t_\rp$.
The components of the matrices $\mathds{N}(k_\rr^+,{k_\rt^-}')$ and $\mathds{L}(k_\rr^+,{k_\rt^-}',{k_\rt^-}'')$ are given by
\begin{align}
  N_{\rs\rs} &=- \hat{\boldsymbol{\kappa}}\cdot\hat{\boldsymbol{\kappa}}' (\gamma_\rt + \gamma_\rt') ,\\ 
  N_{\rs\rp} &= -\frac{\gamma_\rt'}{\sqrt{\epsilon} k_0}(\hat{\boldsymbol{\kappa}}\times\hat{\boldsymbol{\kappa}}')_z (\gamma_\rt + \gamma_\rt'),\\ 
  N_{\rp\rs} &=-\frac{(\hat{\boldsymbol{\kappa}}\times\hat{\boldsymbol{\kappa}}')_z}{\sqrt{\epsilon} k_0}(\gamma_\rr^2 \epsilon + \gamma_\rt \gamma_\rt') ,\\ 
  N_{\rp\rp} &= \frac{1}{\epsilon k_0^2} \bigl[ - \kappa \kappa' (\gamma_\rt + \gamma_\rt' \epsilon) + \hat{\boldsymbol{\kappa}}\cdot\hat{\boldsymbol{\kappa}}' \gamma_\rt' (\gamma_\rr^2 \epsilon + \gamma_\rt \gamma_\rt') \bigr]. 
\end{align}
and
\begin{align}
  L_{\rs\rs} &= (\hat{\boldsymbol{\kappa}}\cdot\hat{\boldsymbol{\kappa}}')
                                                     (\hat{\boldsymbol{\kappa}}'\cdot\hat{\boldsymbol{\kappa}}'') \nonumber\\
                                                  &\quad -\frac{\gamma_\rr' \gamma_\rt'}{{\kappa'}^2 + \gamma_\rr' \gamma_\rt'}
                                                      (\hat{\boldsymbol{\kappa}}\times\hat{\boldsymbol{\kappa}}')_z
                                                      (\hat{\boldsymbol{\kappa}}'\times\hat{\boldsymbol{\kappa}}'')_z ,\\
  L_{\rs\rp} &= - \frac{\gamma_\rt''}{\sqrt{\epsilon} k_0}
                                                      (\hat{\boldsymbol{\kappa}}'\times\hat{\boldsymbol{\kappa}}'')_z 
                                                      (\hat{\boldsymbol{\kappa}}\cdot\hat{\boldsymbol{\kappa}}') \nonumber\\
                                                  &\quad-\frac{\gamma_\rt'}{\sqrt{\epsilon} k_0} 
                                                      (\hat{\boldsymbol{\kappa}}\times\hat{\boldsymbol{\kappa}}')_z 
                                                      \frac{\kappa'\kappa'' + \gamma_\rr'\gamma_\rt'' \hat{\boldsymbol{\kappa}}'\cdot\hat{\boldsymbol{\kappa}}''}{{\kappa'}^2 + \gamma_\rr' \gamma_\rt'} ,\\
  L_{\rp\rs} &=  -\frac{\gamma_\rt}{\sqrt{\epsilon} k_0}
                                                      (\hat{\boldsymbol{\kappa}}\times\hat{\boldsymbol{\kappa}}')_z 
                                                      (\hat{\boldsymbol{\kappa}}'\cdot\hat{\boldsymbol{\kappa}}'') \nonumber\\
                                                  &\quad+\frac{\gamma_\rt'}{\sqrt{\epsilon} k_0} 
                                                      (\hat{\boldsymbol{\kappa}}'\times\hat{\boldsymbol{\kappa}}'')_z 
                                                      \frac{\kappa\kappa'\epsilon -  \gamma_\rt\gamma_\rt'\hat{\boldsymbol{\kappa}}\cdot\hat{\boldsymbol{\kappa}}'}{{\kappa'}^2 + \gamma_\rr' \gamma_\rt'} ,\\
  L_{\rp\rp} &= \frac{\gamma_\rt'\gamma_\rt''}{\epsilon k_0^2}
                                                     (\hat{\boldsymbol{\kappa}}\times\hat{\boldsymbol{\kappa}}')_z
                                                     (\hat{\boldsymbol{\kappa}}'\times\hat{\boldsymbol{\kappa}}'')_z \nonumber\\
                                                  &\quad+\frac{1}{\sqrt{\epsilon} k_0^2}
                                                     \frac{\kappa' \kappa'' + \hat{\boldsymbol{\kappa}}'\cdot\hat{\boldsymbol{\kappa}}'' \gamma_\rr' \gamma_\rt''}{{\kappa'}^2 + \gamma_\rr' \gamma_\rt'}  
                                                  (\kappa\kappa' \sqrt{\epsilon} - \gamma_\rt \gamma_\rt'\hat{\boldsymbol{\kappa}}\cdot\hat{\boldsymbol{\kappa}}').
\end{align}

\section{Definition of the proper self-energy}
\label{Append:DefPropSEnergy}

Inserting the second-order Green's function in Eq.~(\ref{Eq:GreensfunctionAppendix}) into Eq.~(\ref{Eq:Def_LDOSev}) and averaging allows for finding the
second-order surface roughness correction to the LDOS in Eq.~(\ref{Eq:LDOS_sec_ord})-(\ref{Eq:GreenSurfaceWave}) when defining the
proper self-energy $M_{\rs/\rp}$ as 
\begin{equation}
\begin{split}
  M_{\rs/\rp} &= \frac{(k_0 \delta)^2 (-\ri) (\epsilon - 1)}{8 \gamma_\rr^2 (D^0_{\rs/\rp})^2}
                \biggl\{ N_{\rs/\rp}^a(\kappa,\kappa) \\
              &\qquad+ 2 \int\!\!\frac{\rd^2 \kappa'}{(2 \pi)^2}\,g(|\boldsymbol{\kappa} - \boldsymbol{\kappa}'|) N_{\rs/\rp}^b(\kappa,\kappa',\kappa) (\gamma_\rt' - \gamma_\rr') \biggr\}
\label{Eq:ProperSelfEnergy}
\end{split}
\end{equation}
with $N_{\rs/\rp}^{a/b}$ given by
\begin{align}
  N_\rs^a (\kappa,\kappa)         &= (t_\rs)^2  N_{\rs\rs}({k_\rr^+},{k_\rt^-}) \\
  N_\rp^a (\kappa,\kappa)         &= (t_\rp)^2  N_{\rp\rp}({k_\rr^+},{k_\rt^-}) h_\rp \\  
  N_\rs^b (\kappa,\kappa',\kappa) &= (t_\rs)^2 L_{\rs\rs}({k_\rr^+},{k_\rt^-}',{k_\rt^-}) \\ 
  N_\rp^b (\kappa,\kappa',\kappa) &= (t_\rp)^2 L_{\rp\rp}({k_\rr^+},{k_\rt^-}',{k_\rt^-}) h_\rp
\end{align}
Now, inserting the above defined components of the Matrices $\mathds{N}$ and $\mathds{L}$ yields
\begin{align}
  M_{\rs} &= - \ri (k_0 \delta)^2 (\epsilon - 1) \gamma_\rt \nonumber\\
          &\qquad + (k_0 \delta)^2 (\epsilon - 1)^2 k_0^2 \int\!\!\!\frac{\rd^2 \kappa'}{(2 \pi)^2} g(|\boldsymbol{\kappa} - \boldsymbol{\kappa}'|) \\
          &\qquad    \biggl[ D_\rs^0 (\kappa') (\hat{\boldsymbol{\kappa}}\cdot\hat{\boldsymbol{\kappa}}')^2  
          + \frac{D_\rp^0(\kappa')}{k_0^2 \epsilon} \gamma_\rr'\gamma_\rt' (\hat{\boldsymbol{\kappa}}\times\hat{\boldsymbol{\kappa}}')^2 \biggr], \nonumber\\ 
  M_{\rp} &= \ri (k_0 \delta)^2  \frac{(\epsilon - 1)}{\epsilon} \gamma_\rt \biggl(1 - \frac{\kappa^2}{k_0^2} \frac{\epsilon + 1}{\epsilon} \biggr) \nonumber \\
          &\qquad + (k_0 \delta)^2 \frac{(\epsilon - 1)^2}{\epsilon^2} \int\!\!\!\frac{\rd^2 \kappa'}{(2 \pi)^2} g(|\boldsymbol{\kappa} - \boldsymbol{\kappa}'|)\\ 
          & \qquad\biggl[ - \gamma_\rt^2 D_\rs^0 (\kappa') (\hat{\boldsymbol{\kappa}}\times\hat{\boldsymbol{\kappa}}')^2 + \frac{D_\rp^0(\kappa')}{k_0^2 \epsilon} (\kappa \kappa' + \hat{\boldsymbol{\kappa}}\cdot\hat{\boldsymbol{\kappa}}'\, \gamma_\rr'\gamma_\rt) \nonumber \\
          &\qquad \qquad \times (\kappa \kappa' \epsilon - \hat{\boldsymbol{\kappa}}\cdot\hat{\boldsymbol{\kappa}}' \, \gamma_\rt'\gamma_\rt) \biggr]. \nonumber 
\end{align}
Finally, these expressions can be further simplified 
by utilising cylindrical coordinates and introducing the modified Bessel functions   
\begin{equation}
  I_n(x) = \frac{1}{2 \pi} \int_0^{2 \pi}\!\!\rd \varphi\, \cos(n \varphi) \re^{-x \cos(\varphi)}.
\end{equation}
we get for s- and p-polarized modes the relation
\begin{equation}
  M_{\rs/\rp} = M_{\rs/\rp,0} + M_{\rs/\rp,1} 
\end{equation}
with
\begin{align}
  M_{\rs,0} &= -  \ri (k_0 \delta)^2\gamma_\rt (\epsilon - 1) , \label{Eq:Ms0appendix} \\
  M_{\rs,1} &= \epsilon V 
               \int_0^\infty\!\!\rd x\, x \,\re^{-x^2} \bigl\{ 
             k_0^2 \epsilon D^0_\rs(2 x a^{-1}) \nonumber \\
            &\qquad\qquad\times   \bigl[ I_0(\kappa a x) + I_2 (\kappa a x) \bigr] \label{Eq:Ms1appendix}\\
            &\quad + D^0_\rp(2 x a^{-1})\gamma_\rr(2 x a^{-1}) \gamma_\rt(2 x a^{-1}) \nonumber \\
            &\qquad\qquad\times \bigl[ I_0(\kappa a x) - I_2 (\kappa a x) \bigr]\bigr\}, \nonumber 
\end{align}
and 
\begin{align}
  M_{\rp,0} &= \ri (k_0 \delta)^2 \gamma_\rt \frac{(\epsilon - 1)}{\epsilon}\biggl( 1 - \frac{\kappa^2}{k_0^2} \frac{\epsilon + 1}{\epsilon} \biggr), \label{Eq:Mp0appendix} \\
  M_{\rp,1} &= V 
                \int_0^\infty\!\!\rd x\, x \, \re^{-x^2} \biggl\{ \frac{2 }{\epsilon k_0^2}   D^0_\rp(2 x a^{-1})\biggl[
                \epsilon \biggl(\frac{2 x \kappa}{a}\biggr)^2 I_0(\kappa a x) \nonumber \\ 
            &\quad  
             + \gamma_\rt(\kappa)\bigl(\gamma_\rr(2 x a^{-1})\epsilon - \gamma_\rt(2 x a^{-1}) \bigr) (2 x a^{-1} \kappa) I_1(\kappa a x) \nonumber \\
            &\quad    - \frac{1}{2}\bigl( I_0 (\kappa a x) + I_2 (\kappa a x) \bigr) \gamma_\rr(2 x a^{-1}) \gamma_\rt(2 x a^{-1}) \gamma_\rt^2(\kappa) \biggr] \nonumber \\
            &\quad - D^0_\rs(2 x a^{-1}) \gamma_\rt^2(\kappa) \bigl[ I_0(\kappa a x) - I_2 (\kappa a x) \bigr] \biggr\}. \label{Eq:Mp1appendix}
\end{align}

\section{Approximation of the proper self-energy for $|k_0 \sqrt{\epsilon} a/2| \ll 1$}
\label{Append:properselfenergyLDA}

Employing $|k_0 \sqrt{\epsilon} a/2| \ll 1$ for $\gamma_\rr,\gamma_\rt$, and $D_{\rs,\rp}^0$ we find for $M_{\rs,1}$ given in Eq.~(\ref{Eq:Ms1appendix})
\begin{equation}
  \frac{M_{\rs,1}}{V} \approx \biggl[ \epsilon^2 k_0^2 \frac{a}{4}\bigl( I_0^0 (\kappa a) + I_2^0(\kappa a) \bigr) 
            - \frac{\epsilon^2}{\epsilon + 1}\frac{2}{a}\bigl( I_0^2(\kappa a) - I_2^2(\kappa a) \bigr) \biggr],
\label{Eq:Ms1}
\end{equation}
where
\begin{equation}
  V = (k_0 \delta)^2 \frac{(\epsilon - 1)^2}{\epsilon^2} \re^{-\frac{\kappa^2 a^2}{4}}
\end{equation}
and
\begin{equation}
  I^m_n(\xi) = \int_0^\infty\!\!\!\rd x\, x^m \re^{-x^2} I_n(\xi x).
\end{equation}

For the p-polarized modes we get
\begin{equation}
\begin{split}
  \frac{M_{\rp,1}}{V} &\approx \biggl\{ \frac{2}{\epsilon k_0^2} \frac{\epsilon}{\epsilon + 1} \frac{a}{2} \biggl[ 
             (\epsilon - 1) \gamma_\rt(\kappa) \frac{4}{a^2} \kappa I_1^2(\kappa a) \\
            &\qquad +\epsilon \biggl( \frac{2 \kappa}{a}\biggr)^2 I_0^2(\kappa a)+ \frac{2}{a} \gamma_\rt^2(\kappa) \bigl(I_0^2(\kappa a) + I_2^2(\kappa a)\bigr) \biggr] \\
            &\qquad- \frac{a}{2} \gamma_\rt^2(\kappa) \bigl(I_0^0 (\kappa a) - I_2^0 (\kappa a)\bigr) \biggr\}.
\end{split}
\label{Eq:Mp1}
\end{equation} 

The integrals of modified Bessel functions can be integrated analytically yielding~\cite{GradshteynRyzhik2007}
\begin{align}
  I^0_n(\xi) &= \frac{\sqrt{\pi}}{2} \re^\frac{\xi^2}{8} I_{\frac{n}{2}}\biggl(\frac{\xi^2}{8}\biggr), \\
  I^2_n(\xi) &=  \frac{\sqrt{\pi}}{2} \re^\frac{\xi^2}{8} \biggl\{ 
                 \biggl(\frac{1}{2} + \frac{\xi^2}{8}  \biggr)I_{\frac{n}{2}}\biggl(\frac{\xi^2}{8}\biggr) \\
             &\qquad    + \frac{\xi^2}{16}\biggl[I_{\frac{n-2}{2}}\biggl(\frac{\xi^2}{8}\biggr) 
                 + I_{\frac{n+2}{2}}\biggl(\frac{\xi^2}{8}\biggr)  \biggr] 
                  \biggr\}.
\end{align}
By means of these relations one can implement both limits $\kappa a \ll 1$ and $\kappa a \gg 1$ by
approximating the modified Bessel function.

\section{Large distance limit ($z \gg a$ and $a \gg \rd_\rs$)}
\label{Append:LargeDistLimit}

With a similar procedure as above, but in the opposite limit $|k_0 \sqrt{\epsilon} a/2| \gg 1$ we find for $z \gg a$ the
lowest order terms of the self energy
\begin{align}
    M_{\rs,1} &\approx (k_0 \delta)^2 \frac{(\epsilon - 1)^2}{\epsilon^2} \frac{2}{\epsilon k_0^2} (-\ri \gamma_\rt^2 k_0 \sqrt{\epsilon}), \\
    M_{\rp,1} &\approx  (k_0 \delta)^2 \frac{(\epsilon - 1)^2}{\sqrt{\epsilon}} \ri k_0.
\end{align}
As is apparent, in that limit the self energy is independent from the correlation length $a$.

%
%

\end{document}